%% file: alp.tex
\documentclass[a4paper,11pt]{article}
\pdfoutput=1 % if your are submitting a pdflatex (i.e. if you have
\usepackage{jheppub} % for details on the use of the package, please
\usepackage{lineno}
\usepackage[T1]{fontenc} % if needed
\newcommand{\egev}{\ensuremath{\,\textrm{GeV}}}

\title{\boldmath Sensitivity of the FCC-ee to decay of
an axion-like particle  into two photons}

\author[a]{G. Polesello}

\affiliation[a]{INFN Sezione di Pavia, Via Bassi 6, 27100 Pavia, Italy}

\emailAdd{giacomo.polesello@cern.ch}

\abstract{The decay of the $Z$ boson into axion-like particles (ALP) at the $Z$-pole run 
of the proposed CERN FCC-ee collider is investigated. 
We perform a detailed study of final states with three photons or with a single 
photon and missing energy, yielding  an evaluation of the accessible values of 
the ALP coupling to the  photon for ALP masses between 0.1 and 85 GeV.
Special attention is paid to the experimental implications of detecting the 
three-photon signature for the ALP mass region below 10~GeV. 
The FCC-ee run will be able to detect the  ALP 
for couplings down to a few $10^{-3}~\mathrm{TeV}^{-1}$ over the considered mass range
in an effective model where the ALP only couples to the U(1) boson of the Standard Model.}
\begin{document} 
\maketitle
\flushbottom

\section{Introduction}\label{sec::introduction}
\input{intro.tex}

\section{The model and its parameters}\label{sec:model}
\input{model.tex}

\section{Signal and background generation}\label{sec:samples}
\input{samples.tex}

\section{Detector simulation}\label{sec:detector}
\input{detsim.tex}

\section{Three-photon  analysis}\label{sec::prompt}
\input{prompt.tex}

\section{Monophoton analysis}\label{sec::mono}
\input{monophoton.tex}

\section{Combined reach and Conclusions}\label{sec::conclusions}
\input{conclusions.tex}

\bibliography{alp-bibliography}% common bib file

\end{document}

%% file: intro.tex
The next generation of high-energy particle colliders is under
active discussion in the particle physics community.
A very attractive option being discussed are $e^+e^-$ circular colliders, such
as the CERN FCC-ee \cite{FCC:2018evy}. These machines will provide access to
a broad range of physics studies, from precision measurements
of the Higgs boson and of Standard Model (SM) parameters, to direct
searches for physics beyond the Standard Model (BSM).

A special strength of the program is the possibility of studying
the production of new particles in a mass range of 100~MeV-100~GeV for 
extremely low values of their coupling to the
$Z$ boson exploiting the very large projected FCC-ee integrated luminosity 
at the $Z$-pole. These very low couplings can lead to long-lived particles 
decaying inside the detector (LLPs), allowing for searches with 
very low Standard Model backgrounds. This requires special care in the
design of the detectors, and detailed studies are needed to understand 
the constraints imposed on the detectors by different final-state 
new physics signatures \cite{Blondel:2022qqo}.

Of particular interest is the production of pseudoscalar axion-like particles 
(ALP, $a$ in formulas), pseudo Nambu-Goldstone bosons which appear in theories 
where an approximate global symmetry of the theory is spontaneously broken.
Such particles feature in several extensions of the Standard Model, and provide 
final-state signatures which can be explored at future colliders. Such signatures are
discussed in detail in~\cite{Mimasu:2014nea, Brivio:2017ije} 
and~\cite{Bauer:2017ris, Bauer:2018uxu}, based on effective models for the
interactions of the ALP with SM particles.

The present study aims at evaluating the sensitivity for the production of 
such particles for the $Z$-pole run of the FCC-ee, 
%at a c.m.s.  energy $\sqrt{s}=91\egev$ and 
with target integrated luminosity 
$L_{int}= 2.05\times 10^8\, \mathrm{pb}^{-1}$, distributed among
the center-of-mass energies $\sqrt{s}=88, 91, 94\egev$, corresponding to the production
of approximately $6\times10^{12}$ $Z$ bosons. The ALP mass region ranging from $0.1$ to 
$90 \egev$ is investigated, based on  the model defined in 
\cite{Bauer:2018uxu,Bauer:2017ris},  
for the process $e^+e^-\rightarrow Z a$ followed by the decay $a\rightarrow\gamma\gamma$. 

Preliminary studies for this process in the long-lived case
are shown in \cite{Blondel:2022qqo} where the FCC-ee reach is 
given as a purely theoretical estimate of the number of produced events
decaying inside a nominal detector volume.
%where  
%the experimental reach is defined as the parameter space for which
%four events decaying inside the detector are produced for the 
%expected FCC-ee Z-poles luminosity.
For the prompt decays, the study \cite{Steinberg_2021} addresses the Tera-Z option of 
the ILC, and it identifies the main experimental issues affecting the search.

The experimental reach of FCC-ee is assessed here based on a parametrised 
simulation of the proposed IDEA detector \cite{Antonello:2020tzq} 
and on the consideration of the leading 
irreducible backgrounds. This approach provides a realistic
estimate of the physics coverage and valuable information on its 
dependence key elements of the detector performance.
In the next paragraphs we will briefly comment on the 
limitations of the study, which will be addressed in the near future 
as the development of the FCC-ee detectors will converge towards the final design.
%A detailed consideration of the geometry of the proposed detector, based
%on a full GEANT4 simulation is needed for this study, and is outside
%the scope of the present study.

Two cases will be studied: in the first case the ALP decays inside the  
detector  yielding a three-photon signature; in the second case the
decays happens outside the detector, and the final state is a single 
photon and missing energy.  For the decays inside the detector no distinction will 
be made between `prompt' ALP decays, happening near the production point,
and long-lived ones.  The latter case requires a detailed study of the pointing 
performance of  the electromagnetic calorimeter, which is not presently available.
Likewise, for masses of the ALP below 1 GeV, and generically for long-lived ALPs,
a good understanding of the shower position measurement and  
of the shower-shower separation of the electromagnetic calorimeter is
needed. In the following, this will be replaced by parametrisations of the 
performance obtained from published simulation work by the calorimeter communities.  
A complete assessment of these performance aspects will require dedicated studies 
based on the GEANT4 simulation \cite{GEANT4:2002zbu} of the full detector with $4\pi$ geometry,
and the development of dedicated reconstruction algorithms.  

Only irreducible backgrounds for the three-photon case are considered. 
Existing LEP studies  give confidence that reducible backgrounds can 
be beaten down to a level well below the irreducible ones 
\cite{L3:1994shn,OPAL:1990tfv,L3:1995nbq,DELPHI:1991emv}. 
However given the extremely high $Z$ statistics, this statement will need
to be backed up by detailed detector studies.

The production of the ALP in the decay of the $Z$ boson is addressed in this
paper. An alternative production process, where the ALP is produced in the 
fusion of two photons radiated by the electron and the positron is studied 
in \cite{RebelloTeles:2023uig}, and it is shown to cover a significant 
and complementary part of the parameter space.

The paper is organised as follows: the effective ALP model is first introduced
and briefly discussed.  In the following sections the Monte Carlo event generation 
is discussed, followed by a description of  the adopted approach to detector simulation. 
On this basis two analyses are  developed in the two following sections, addressing 
respectively the three-photon and the one-photon final state. Finally the additional 
coverage of FCC-ee with respect to existing searches is assessed for the studied 
benchmark model.

%% file: model.tex
The present study  is based on the ALP effective model described in \cite{Bauer:2018uxu,Bauer:2017ris}, which
was also used in the previous FCC-ee study documented in \cite{Blondel:2022qqo}.

\begin{figure}[h!]
\centering
\includegraphics[width=0.45\textwidth]{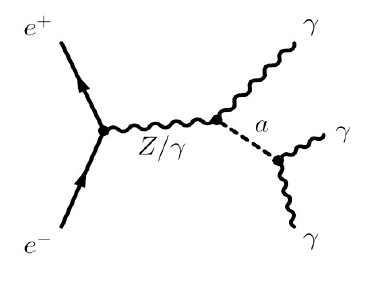}
\caption{Feynman diagram for production and decay of the ALP at FCC-ee.}\label{fig:alpprod}
\end{figure}

The channel of interest for the $Z$-pole study is the decay of the $Z$ boson into
a photon and an ALP, 
%whose strength is defined by the coupling $C_{\gamma Z}$,
followed by the decay \mbox{$a\rightarrow\gamma\gamma$}, yielding a three-photon
final state, shown in Fig.~\ref{fig:alpprod}.

The part of the Lagrangian relevant to the production and decay of interest  describes the 
coupling of  the  ALP  to the photon and the $Z$ boson after electroweak symmetry breaking as \cite{Bauer:2018uxu}:
\begin{equation}\label{gammaZ}
   {\cal L}_{\rm eff}
   \ni e^2\,C_{\gamma\gamma}\,\frac{a}{\Lambda}\,F_{\mu\nu}\,\tilde F^{\mu\nu}
    + \frac{2e^2}{s_w c_w}\,C_{\gamma Z}\,\frac{a}{\Lambda}\,F_{\mu\nu}\,\tilde Z^{\mu\nu}
    + \frac{e^2}{s_w^2 c_w^2}\,C_{ZZ}\,\frac{a}{\Lambda}\,Z_{\mu\nu}\,\tilde Z^{\mu\nu} \,.
\end{equation}
where $s_w$ and $c_w$ are the sine and cosine of the weak mixing angle, $\Lambda$ the new physics scale, and $F_{\mu\nu}$ and   $Z^{\mu\nu}$ describe the photon, and $Z$ boson 
in the broken phase of EW symmetry.  The relevant Wilson coefficients 
$C_{\gamma\gamma}$, $C_{\gamma Z}$ and $C_{ZZ}$ can be written as:
\begin{equation}\label{WilsonCoeff}
C_{\gamma\gamma}=C_{WW}+C_{BB}, \hspace{1cm} 
C_{\gamma Z}=c^2_w\,C_{WW}-s^2_w\,C_{BB}, \hspace{1cm} 
C_{ZZ}= c^4_w\,C_{WW}+ s^4_w\,C_{BB}\,,
\end{equation}
with $C_{WW}$ and $C_{BB}$  the Wilson coefficients for the coupling of the ALP
to the unbroken SU(2) and U(1) gauge fields.

Following \cite{Bauer:2018uxu} we adopt a benchmark model where the tree-level
couplings of the ALP to fermions and gluons are set to zero, and the ALP only
couples to the U(1) gauge fields, i.e., the coefficient $C_{WW}$ is set to zero. 
In this situation, from the formulas \ref{WilsonCoeff}, 
$C_{\gamma Z}=-s^2_w\,C_{\gamma\gamma}$, and the model is described by only two 
two parameters, the ALP mass $m_a$ and $C_{\gamma\gamma}$.
The branching ratio \mbox{$BR(a\rightarrow\gamma\gamma)$} is 100\%, and for  
$C_{\gamma\gamma}=1$, and $\Lambda=1$~TeV the cross-section at the $Z$ pole is $2.7$~pb for 
$m_{a}=1$~GeV.

As only the $\gamma\gamma$ decay is open, the total width of the ALP is 
\begin{equation}
 \Gamma(a)  = \frac{4\pi\alpha^2 m_a^3}{\Lambda^2}\,\big| C_{\gamma\gamma}^\text{eff} \big|^2
\end{equation}
For sufficiently low values of $m_a$ and $G_{\gamma\gamma}$ the  ALP is long-lived,  
i.e. it has a measurable decay length $L_a$ in the detector.
As a guide to the experimental studies, it is useful to evaluate in the
parameter space of the model how many ALP decays into two photons are expected 
for a statistic of $6\times10^{12}$ $Z$-bosons, and in what region of this space
they can be long-lived.
The result is illustrated in Figure~\ref{fig:lines_alp}, and is classified in four regions:
\begin{figure}
\centering
\includegraphics[width=0.75\textwidth]{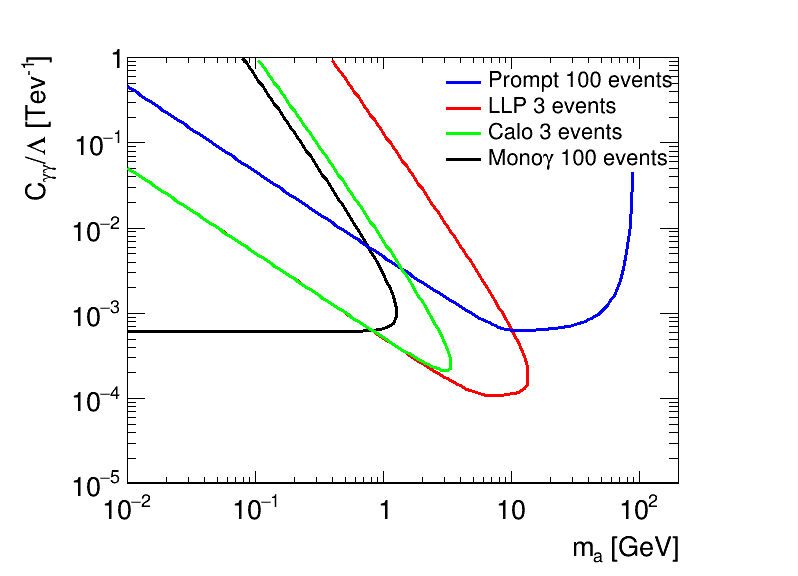}
\caption{Areas in parameter space which can be addressed by different kinds of
experimental ALP searches at the FCC-ee $Z$-pole run for the process $\rm e^+ e^- \to \gamma a \to 3\gamma$.} \label{fig:lines_alp}
\end{figure}
\begin{itemize}
\item
The `Prompt' region, with $L_a<1$~cm, where the path of the ALP in the detector
does not affect the measurement of the kinematic properties. 
Since a large irreducible background is present for this case, consisting
in the QED $e^+e^-\rightarrow\gamma\gamma\gamma$ process, the line bounding the
region where at least 100 events are produced for the FCC-ee Z-pole run
is shown.
\item
The `LLP' region, where the ALP decays before reaching the calorimeter,
and its path is in principle measurable, corresponding to 
$1~\mathrm{cm}<L_a<200~\mathrm{cm}$ for the geometry 
of the IDEA detector. In this case no irreducible background is expected, and 
the curve for three events is shown.
\item
The `Calo' region where the ALP decays inside the calorimeter,
corresponding to $200~\mathrm{cm}<L_a<400~\mathrm{cm}$. The signature
is a prompt monochromatic photon recoiling against an energy deposition
in the calorimeter. If it can be experimentally shown that the
shower starts beyond the expected starting point of an electromagnetic 
shower in the calorimeter, this signal could also have no irreducible
background, and the curve for three events is shown.
\item	
The `Mono$\gamma$' region where the ALP decays outside the detector, yielding
the signature of a prompt monochromatic photon recoiling against missing momentum.
In this case the decay length requirement is $L_a>450$~cm. The irreducible background
for this channel is $e^+e^-\rightarrow \gamma Z, Z\rightarrow \nu\nu$, and the
curve for 100 events is shown.
\end{itemize}
From Figure~\ref{fig:lines_alp}, the different approaches cover complementary
areas in parameter space. The separation between `Prompt' and `LLP'
relies on the experimental measurement of the impact angle of the photons on
the calorimeter, which can be used to determine the position of ALP decay in the detector.
This has been shown to be possible by the ATLAS Collaboration 
\cite{ATLAS:2014kbb,ATLAS:2022vhr}, but,
as discussed in the introduction, no study on the feasibility of this measurement
and on the achievable measurement precision is available yet for the proposed FCC detectors.
We will therefore develop a three-photon analysis where the ALP is required to decay
at truth level inside the central detector, modelled as cylinder with 
radius 2~m and length 4~m, and no attempt is made to exploit the measurement
of the impact angle of the detected photons. This analysis should cover 
both the `Prompt' and the `LLP' regions, and it will referred to in the
following as three-photon analysis.
In addition we will perform a single-photon analysis addressing the `Mono$\gamma$' region.

%% file: samples.tex
%\section{Signal and background generation}
The signal samples were generated with \verb+MG5aMC@NLO+ \cite{Alwall:2014hca}, based on the 
\verb+ALP_NLO_UFO UFO+ \cite{Bauer:2018uxu}. 
All the ALP couplings are set to zero except $C_{BB}$. 
This choice of parameters yields $C_{\gamma\gamma}=C_{BB}$,  
and a value of $C_{\gamma Z}=-s^2_wC_{\gamma\gamma}$ which determines the production cross-section for the process of interest,
$$
e^+e^-\rightarrow Z, Z\rightarrow \gamma a
$$
%The $f_{alp}$ parameter is set to 6.33 GeV, corresponding
%a value of 1~TeV ($\Lambda=f_{alp}\times16\times\pi^{2}$) for the $\Lambda$ suppression factor of the couplings.
The $\Lambda$ suppression factor was set at 1~TeV.
A scan was performed over the mass of ALP between 0.1 and 85 GeV, with  
$C_{\gamma\gamma}=1$, and no lifetime simulation for the ALP. A total of 100k events 
per point were generated for these samples, which were used to study the kinematic
selection for the prompt analysis.  For masses below 10~GeV additional samples with simulation of the ALP lifetime for $C_{\gamma\gamma}$ between $5\times10^{-4}$  and $5\times10^{-2}$ 
and lifetime simulation were generated and used to study the impact of long ALP lifetimes 
on the analysis.

The LHE files were hadronised with PYTHIA8 \cite{Sjostrand:2014zea} and then fed into 
the DELPHES \cite{deFavereau:2013fsa} fast simulation of the IDEA 
Detector \cite{Antonello:2020tzq}, based on the official datacards used for the 
"Winter2023" production of the FCC-PED study \cite{winter2023setup}.

%For  the backgrounds from $Z$ decays, the official samples produced
%by the central software group for the FCC under the tag
%"Winter2023" were used \cite{winter2023samples}.

For the prompt analysis the irreducible background from the process 
$$
e^+e^-\rightarrow \gamma \gamma \gamma 
$$
was produced at LO with MG5aMC@NLO.
The only generation-level  requirements were 
that photons be produced within a pseudorapidity $\eta$ of $\pm2.6$.
The cross-section for the process is 9.5~pb, and a sample of 10M events
was produced, corresponding to approximately 1/200 of the expected statistics 
for the FCC-ee run at the $Z$ pole. The events were then processed through 
the same PYTHIA8-DELPHES chain as the signal events. 

For the monophoton analysis two background samples were likewise 
produced at LO with MG5aMC@NLO and processed through
the PYTHIA8-DELPHES chain:
$$
e^+e^-\rightarrow \gamma\nu\nu
$$
and
$$
e^+e^-\rightarrow \gamma e^+e^-
$$
In both cases the matrix-element photon is required to be within
$|\eta|<2.6$ and to have an energy $E_{\gamma}$ in a range around 45.5 GeV,
which is the energy of the monochromatic photon for the ALP masses
below 5~GeV, of interest for the monophoton
signature. For the $\gamma\nu\nu$ samples two ranges were generated:
$E_{\gamma} \in (44.5-46.5)$~GeV for a cross-section of 0.095~fb and 
$E_{\gamma} \in (40-50.5)$~GeV for a cross-section of 1.7~fb. For each sample 1M events were
generated. The two samples are designed to provide adequate MC
statistics for photon energy ranges 
compatible with the resolutions respectively of the crystal and fibre
electromagnetic  calorimeters. For the $\gamma e^+e^-$ sample the photon energy was
required to be in the $40-50.5$~GeV range, and the electrons were required
to have pseudorapidity $|\eta|>2.7$, yielding a cross-section of 0.016~pb.
A million events were generated as well for this sample.

%% file: detsim.tex
The signatures under considerations involve events
with only photons in the final state. Various aspects
of the  performance of the electromagnetic (EM) calorimeter determine
the sensitivity of the analysis. 

In particular, for ALP masses below $\sim10$~GeV, the detection of the ALP decays 
depends on the precision with which the impact of each photon 
in the calorimeter can be measured. Moreover for low ALP masses,
below $m_{a}=1$~GeV the two photons for the ALP decay can be 
very collimated. The photon-photon separation power of the 
EM calorimeter, as well as the precision in mass reconstruction for 
two very collimated photons are essential performance figures for 
the ALP analyses. To model these effects, the DELPHES simulation
was supplemented with dedicated parametrisations for the EM calorimeters,
based on results from  the detailed GEANT4 simulation
of the calorimeter systems being proposed for the IDEA detector.

Two options are presently being considered for the electromagnetic 
calorimeter of IDEA,  one based on a dual readout fibre calorimeter, and one based
on scintillating crystals, with the latter being the baseline option.

For both options a geometry is assumed where the face of the calorimeter
is modelled as a cylinder with 2~m radius and 4~m length, with the 
interaction point and origin of the coordinate system on the
cylinder axis halfway between the two cylinder ends.

For the fibre option, the calorimeter is built as a matrix
of steel tubes with 2~mm diameter, and inside each of them is located
a fibre with 1~mm diameter. Each fibre is read out by a single SiPM,
yielding lateral sampling of the electromagnetic shower with a granularity
of 2~mm. No longitudinal segmentation is foreseen, but the arrival time
of the signal to the SiPM can be measured and used to study the longitudinal
shower development. The performances of a prototype module with this 
geometry have been studied with a GEANT4 simulation. The most updated
results are shown in recent talks by the HIDRA collaboration
\cite{Pareti:2024nnv,Centeno:2024uzn}, and are ($E$ in GeV, $x$,$y$ in mm):
\begin{itemize}
\item
For the energy resolution:
$$
\frac{\sigma(E)}{E}=\frac{0.139}{\sqrt{E}}\,+\,0.006
$$
\item
For the position resolution (in mm):
$$
\sigma(x) =\frac{4.05}{\sqrt{E}}\,+\,0.0;\;\;\;\sigma(y)=\frac{3.23}{\sqrt{E}}\,+\,0.0055
$$
\end{itemize}

For the crystal option \cite{Lucchini:2022vss}, the front face of the crystals 
has a nominal size of $1\times1$~cm, and the calorimeter is segmented 
in depth with a front crystal with nominal length 5~cm, and a rear crystal
with nominal length 15~cm \cite{Chung_CALOR}. The following performance figures are assumed
($E$ in GeV, $\theta$ in mm)
\begin{itemize}
\item
For the energy resolution the value is taken from the DELPHES IDEA card,
and is:
$$
\frac{\sigma(E)}{E}=\frac{0.03}{\sqrt{E}}\,\oplus\,0.005\,\oplus\,\frac{0.002}{E}. 
$$
\item
The position resolution is quoted in \cite{Lucchini:2022vss} as:
$$
\sigma(\theta)=\frac{1.5}{\sqrt{E}}\,\oplus\,0.33;
$$
when transformed in length coordinates on the cylinder surface, this 
yields a value similar to the one for the fibre calorimeter, but with a significantly 
higher constant term.
\end{itemize}

The simulation procedure starts from the selection of the truth photons produced
by PYTHIA within $|\eta|<2.6$, with the starting point inside the inner detector,
and with an energy in excess of 100~MeV.

The $\theta$ and $\phi$ values of the intersection point of the 
momentum vector of the photons with the cylinder representing 
the calorimeter face are calculated, yielding the photon 
impact point in cylindrical coordinates. Thereafter the energy and the 
position of the impact point are smeared according to the resolution 
values given above.

\begin{figure}
\centering
\includegraphics[width=0.45\textwidth]{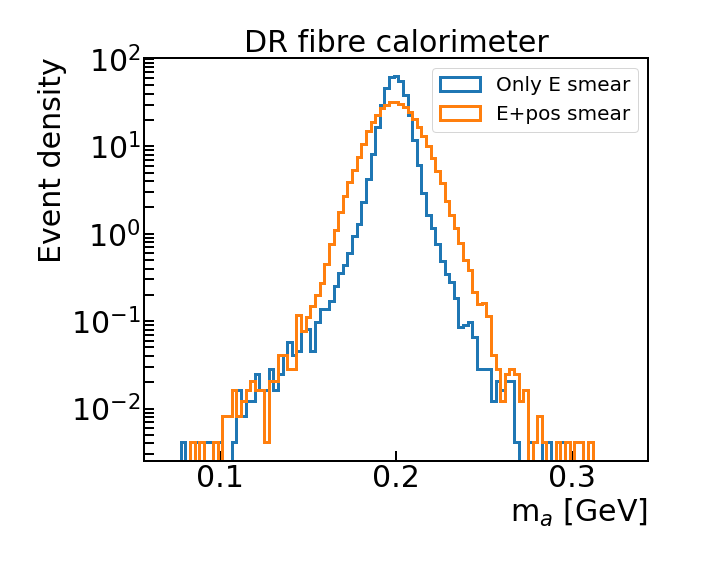}
\includegraphics[width=0.45\textwidth]{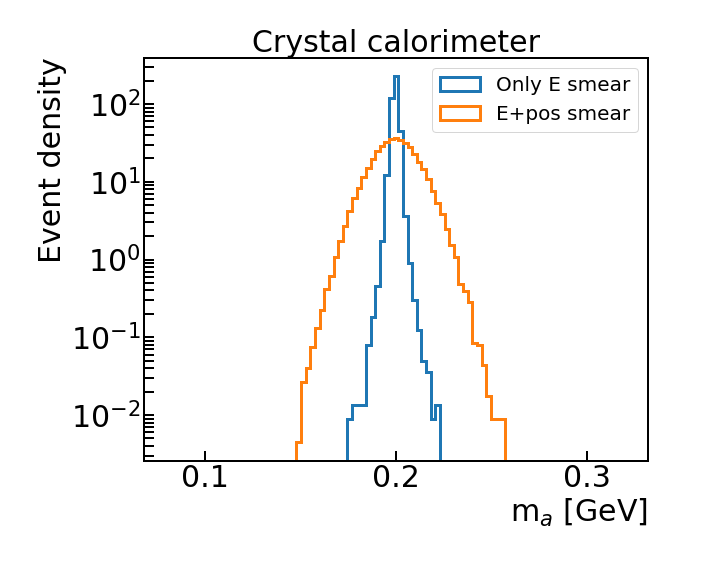}
\includegraphics[width=0.45\textwidth]{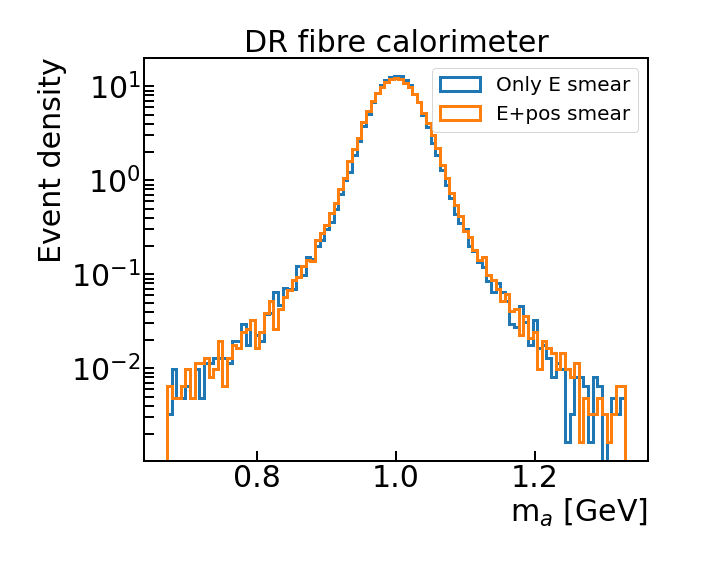}
\includegraphics[width=0.45\textwidth]{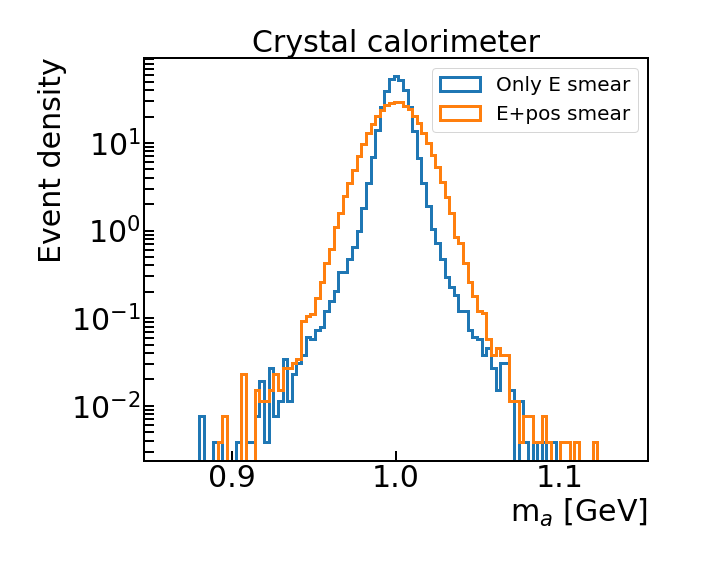}
\includegraphics[width=0.45\textwidth]{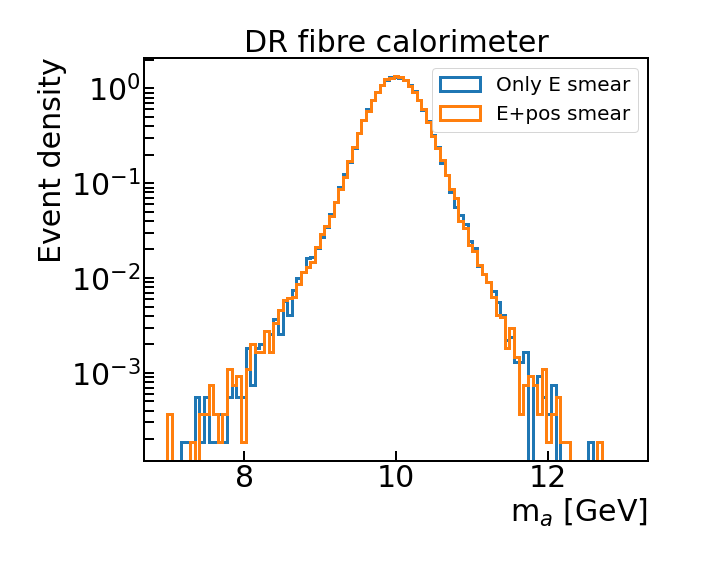}
\includegraphics[width=0.45\textwidth]{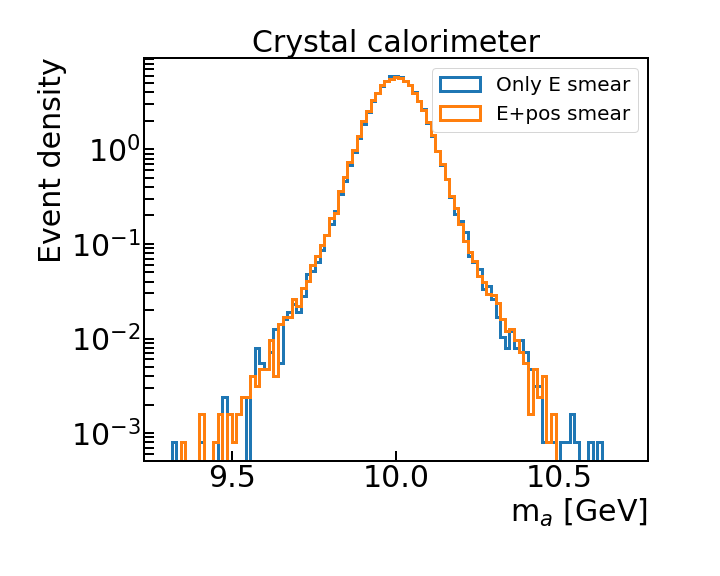}
	\caption{Reconstructed ALP mass for three different test masses: 0.2~GeV (top), 
	1~GeV (center) and 10~GeV (bottom). In the left (right) column  the results 
	for the fibre (crystal) EM calorimeters are shown. The blue line
	shows the resolution when only energy smearing is applied to the photons,
	the orange line when both energy and position smearing are applied.} \label{fig:mass_res}
\end{figure}

In Figure~\ref{fig:mass_res} the resolution of
the measured ALP mass is shown for the two different calorimeter designs.
The results are shown for three ALP test masses, 0.2, 1 and 10 GeV, 
and in each plot the resolution obtained smearing both energy and impact
position of the photon is compared to the one obtained smearing only the
energy.  At the very lowest mass considered the resolution is dominated
by the impact point resolution, and is approximately the same for the 
two calorimeters. The contribution of the position resolution decreases 
for increasing masses, and is approximately negligible at 10 GeV.

Two additional issues affect the experimental sensitivity at low masses:
the experimental merging of the two photons from the ALP decay, 
and the fact that, as discussed in the previous section, in part of the
parameter space the ALP will have a measurable path in the detector 
before decaying.

The first issue is discussed in detail in \cite{Steinberg_2021}
in the framework of an ILC analysis. The minimum distance in radians between two 
photons from the decay
of a resonance of mass $m_{a}$ and energy $E_{a}$ is given by the formula
$$
\Delta \alpha(\gamma\gamma)=2\,m_{a}/E_{a},
$$
yielding values of 0.004, 0.02, 0.04 radians for 0.1, 0.5 and 1 GeV respectively,
which at a distance of 2~m from the decay point correspond to 0.8, 40 and 80~mm.
If the distance between two showers is smaller than the Moliére radius of the calorimeter, 
the two showers are not fully separated, which may mean that the two 
photons are seen as one, or that there is added uncertainty 
on the mass measurement from the uncertainty on the energy sharing between 
the two clusters.

Detailed simulations incorporating the detailed geometry of the
calorimeter systems are needed to understand these effects. Thanks to the high granularity 
and the small Moliere radius (2.2 and 2.4~cm respectively for crystal and fibre)
of the calorimeters under consideration, modern imaging techniques based 
on convolutional neural networks should allow photon-photon mass
reconstruction down to values of $m_{a}$ of $\sim0.1~\mathrm{GeV}$.
This is demonstrated in an existing  study based on the CMS calorimeter
\cite{CMS:2022wjj}. In that case a visible mass peak is obtained for 
$\gamma\gamma$ resonances with masses as low as 100~MeV, and transverse
momenta in the range 30-55~GeV. Considering the higher granularity of the 
calorimeters under discussion for FCC, and the larger distance of
the calorimeters from the interaction point, one can expect even better
performance in the FCC case. For the present study, a sharp cutoff on 
the angular distance between the two photons from the ALP decay $\Delta\alpha$ 
is applied, and events with $\Delta\alpha$ below a given value are rejected in the 
analysis. The considered values of $\Delta\alpha$ are 0.01, 0.02, 0.03 radians,
corresponding respectively to a distance between the two photons
on the face of the calorimeter of respectively 2, 4 and 6~cm. 
The default value for the crystal calorimeter is taken as 0.02 radians, corresponding
to four cells, and for the fibre calorimeter as 0.01 radians, corresponding to $\sim10$ fibres.

\begin{figure}
\centering
\includegraphics[width=0.6\textwidth]{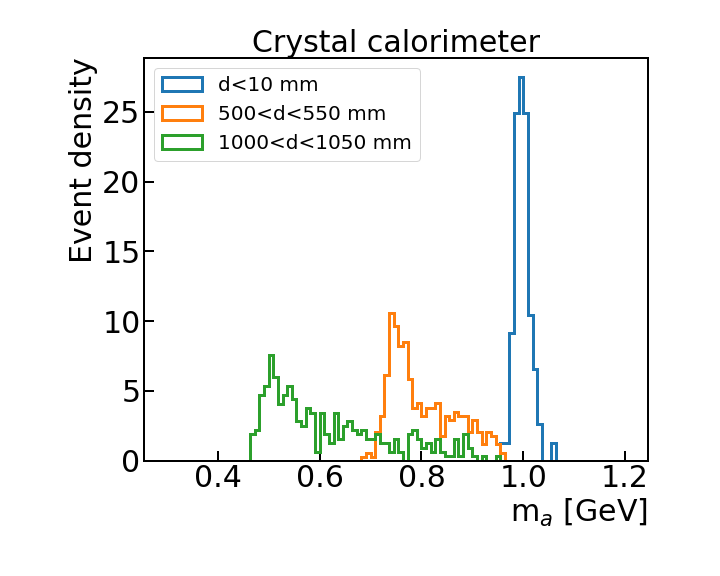}
\caption{Reconstructed ALP mass for $m_{a}=1$~GeV, and three different bins
in the distance of the reconstructed vertex from the center of the
detector, around zero (blue), 500 (orange) and 1000 (green) ~mm respectively.
The crystal calorimeter configuration is shown.} \label{fig:massver}
\end{figure}

The second effect to be considered is the impact of the long lifetime
of the ALP on the measurement of the kinematics.
The mass reconstruction algorithm assumes that the ALP decays in the center of the 
detector. The the wrong angle between the two photons will thus be used in the 
invariant mass calculation, biasing the reconstruction of the ALP mass peak.
%This effect could be partially corrected for if experimental information on the
%impact angle of the photons to the calorimeter is available.
The parametrised simulation described above describes this 
effect, as shown in Figure~\ref{fig:massver}, where the reconstructed 
value of $m_a$ is plotted for $m_{a}=1$~GeV, and three different bins 
in the distance of the reconstructed vertex from the center of the
detector, around zero, 500 and 1000~mm respectively. As expected, for 
the displaced vertexes the mass is reconstructed incorrectly, thus affecting
the kinematic selections.

%% file: prompt.tex
The final state for the process of interest involves three photons
of which one is produced in the decay of the $Z$ ($\gamma_3$) in the
following, and two ($\gamma_1$ and $\gamma_2$) are produced 
in the decay of the ALP. The first step of the analysis consists 
in assigning each of the detected photons to one of the two classes.

If a specific test mass of the ALP, $m_{a}$ is considered, the 
energy of the photon from the $Z$ decay has a fixed energy $E$, determined
by the recoil formula:
\begin{equation}\label{eq:recoil}
E=\frac{E_{CM}^2-m_{a}^2}{2\,E_{CM}}
\end{equation}
Where $E_{CM}$ is the center-of-mass-energy in of the collisions,
91.2 GeV for the present study.

Assume a test mass $m_{a}$, and the corresponding value of the
recoiling photon $E$. For each event and for each of the possible three
assignments of one of the detected photons as $\gamma_3$,
the  variable $M_{cut}$ can be built as:
%which measures the compatibility of that assignment
%for that event as:
\begin{equation}
	M^2_{cut}=\frac{(m_{\gamma_1\gamma_2}-m_{a})^2}{\sigma(m_{a})^2}+\frac{(E_{\gamma_3}-E)^2}{\sigma(E)^2}
\end{equation}
Where the $\sigma(m_{a})$ and $\sigma(E)$ are respectively the expected 
photon-photon mass resolution, and photon energy resolution. These values depend on the
detector configuration and  can be calculated from the simulation.
The variable $M_{cut}$ measures the compatibility of a given assignment with the
expected signal kinematics.

The assignment which yields the minimum value for $M_{cut}$ is taken
as the correct one for the test mass $m_{a}$, and all the variables used in the
analysis are calculated based on this assignment.

\begin{figure}
\centering
\includegraphics[width=0.5\textwidth]{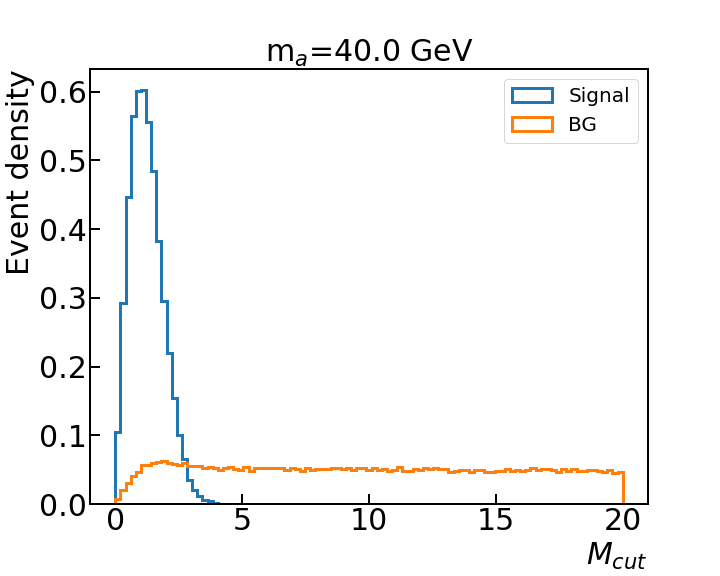}
	\caption{Distribution of the variable $M_{cut}$ for
	signal (blue) and background (orange) for $m_{a}=40$ GeV.}\label{fig:sigmam}
\end{figure}

The variable $M_{cut}$ is approximately
independent of $m_{a}$, and of the studied calorimeter
configuration, and has a good signal to background separation power.
The distribution of $M_{cut}$, truncated at a value of 10 
is shown for $m_{a}=40$~Gev in Figure~\ref{fig:sigmam}, 
for both signal and background. 
%A request of $\sigma(m_{test})<2.5-3$ 
%keeps most of the signal and rejects a large fraction of the background.
Besides the selection on $M_{cut}$, additional separation power
between signal and background can be obtained through a detailed study
of the angular distributions of the final state photons.

\begin{figure}
\centering
\includegraphics[width=0.45\textwidth]{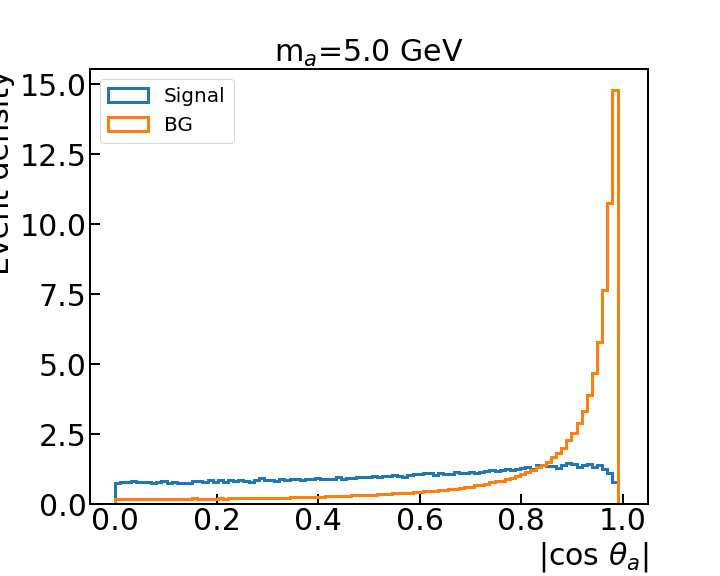}
\includegraphics[width=0.45\textwidth]{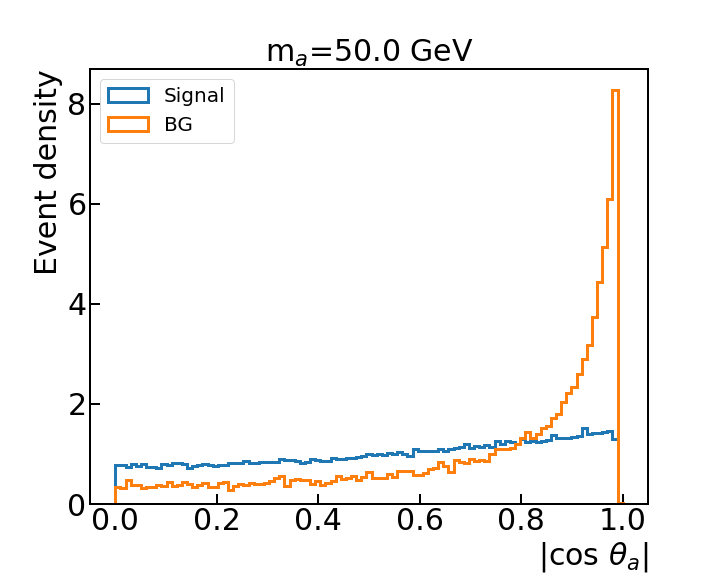}
\caption{Distribution of the variable $\cos\theta_{ALP}$ for
	signal (blue) and background (orange) for two values of $m_{a}$, 5 (left) and 50 (right) GeV.} \label{fig:cthalp}
\end{figure}

\begin{figure}
\centering
\includegraphics[width=0.45\textwidth]{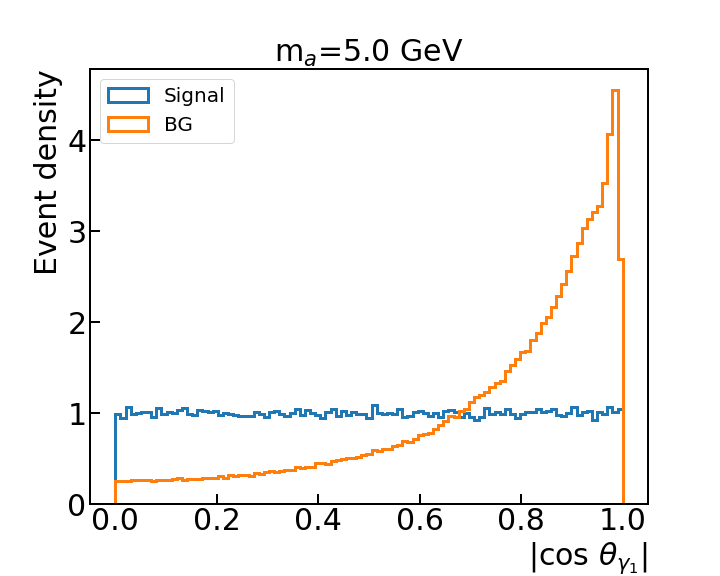}
\includegraphics[width=0.45\textwidth]{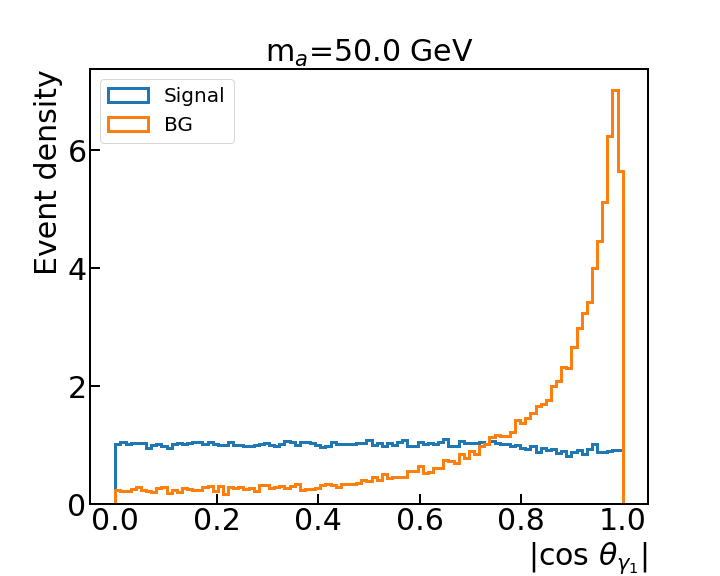}
\caption{Distribution of the variable $\cos\theta_{\gamma_1}$ for
        signal (blue) and background (orange) for two values of $m_{a}$, 5 (left) and 50 (right) GeV.} \label{fig:cthg1}
\end{figure}

\begin{figure}
\centering
\includegraphics[width=0.45\textwidth]{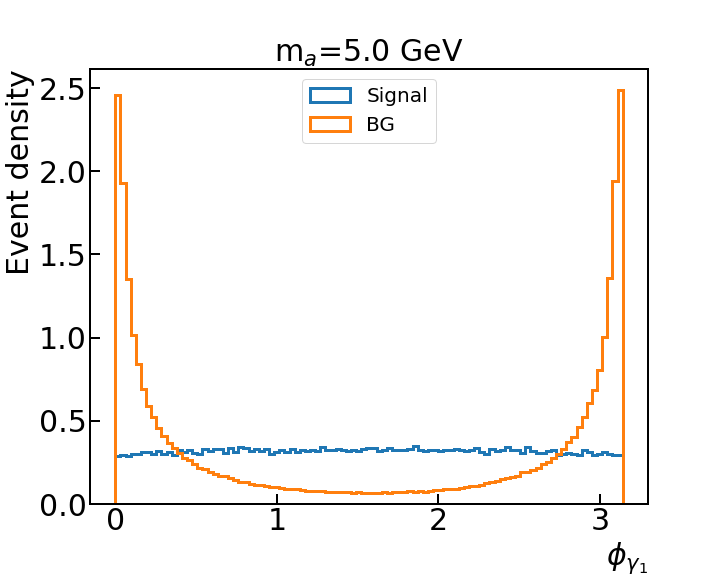}
\includegraphics[width=0.45\textwidth]{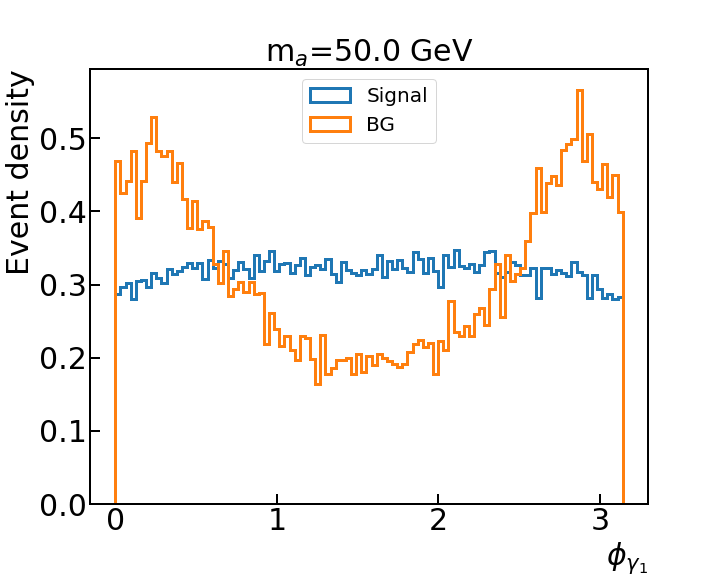}
\caption{Distribution of the variable $\phi_{\gamma_1}$  calculated 
	in the rest frame of the ALP for
        signal (blue) and background (orange) for two values of $m_{a}$, 5 (left) and 50 (right) GeV.} \label{fig:phig1}
\end{figure}

Each event is fully defined by 9 variables, i.e. the three-momenta of the
photons. The energy and momentum constraints from the $e^+e^-$ collision
reduce the independent variables to five.
The signal is a sequence of two-body decays, therefore a convenient choice
of variables is the ALP mass built as the invariant mass of 
$\gamma_1$ and $\gamma_2$, the $\theta$ and $\phi$ of the 
photon from the $Z\rightarrow\gamma a$ decay in the lab frame, 
and $\theta$ and $\phi$ of one of the photons from the $a\rightarrow\gamma\gamma$ decay, 
calculated in the rest frame of the ALP. Given the cylindrical symmetry of the 
system around the beam axis, the events  can be rotated in such a way 
that $\phi_{\gamma_3}=0$.  After this rotation, the event is fully defined by
$m_{\gamma_1\gamma_2}$ and by three angular variables. For the event selection
we use $\cos\,\theta_{\gamma_3}$=$-\cos\,\theta_{ALP}$, 
$\cos\,\theta_{\gamma_1}$ and $\phi_{\gamma_1}$ where the latter two are calculated
in the ALP rest frame.
The distributions of the three angular variables are shown in 
Figures~\ref{fig:cthalp}-\ref{fig:phig1} for two different test masses and for both 
signal and background.
\begin{figure}
\centering
\includegraphics[width=0.5\textwidth]{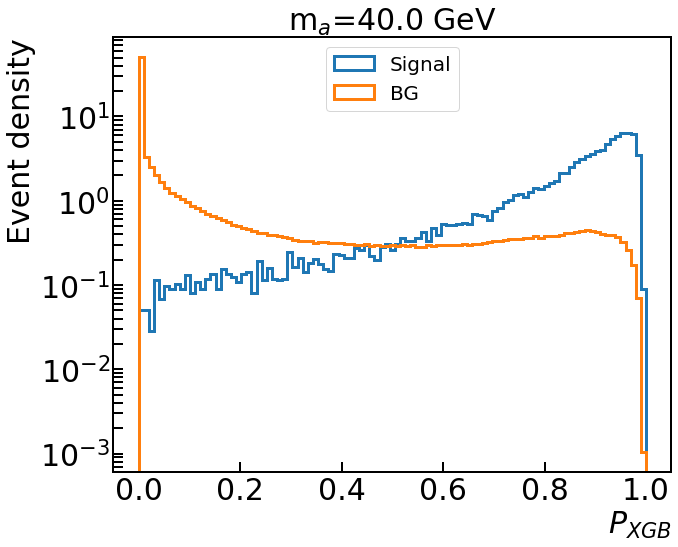}
	\caption{Distribution of the variable $P_{XGB}$ for
signal (blue) and background (orange) for $m_{a}=40$ GeV.}\label{fig:XGB}
\end{figure}

\begin{figure}
\centering
\includegraphics[width=0.495\textwidth]{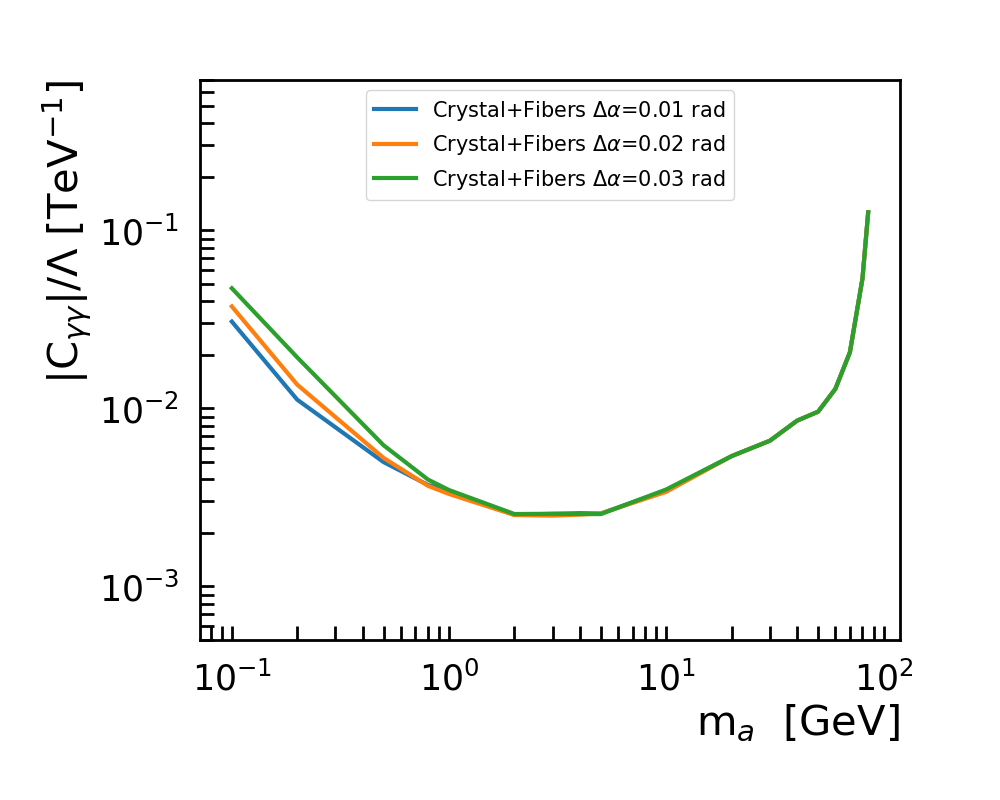}
\includegraphics[width=0.495\textwidth]{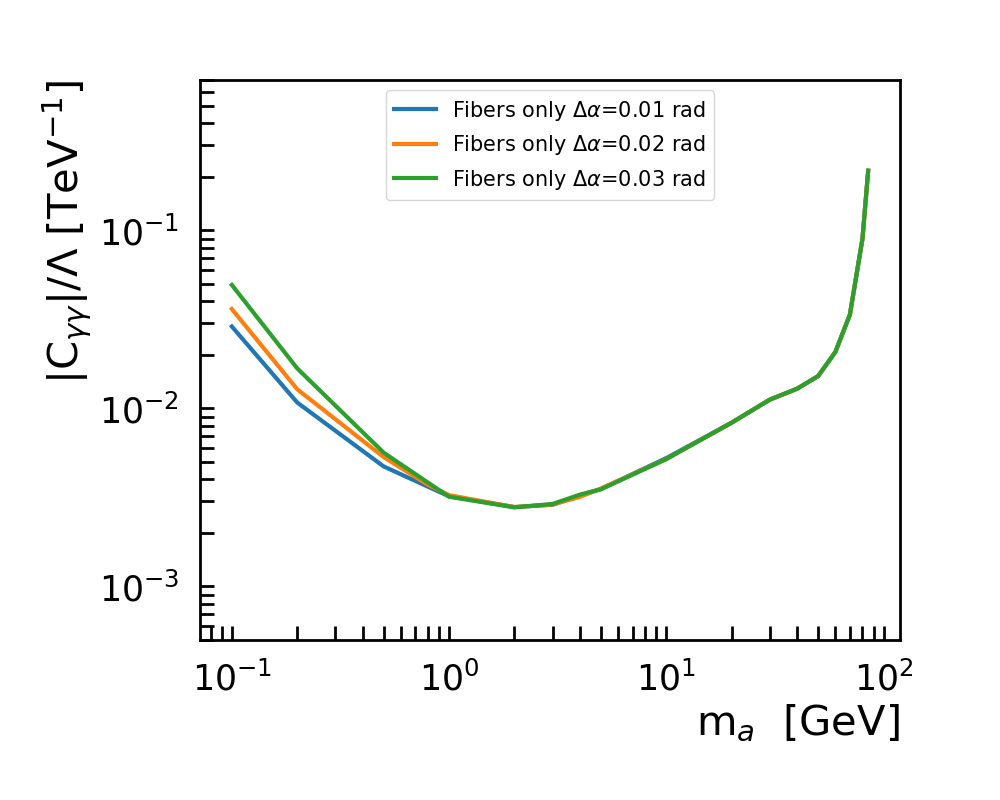}
\caption{95\% CL sensitivity for the $C_{\gamma\gamma}$ coupling as a function of $m_{a}$ for the crystal
(left) and fibre (right) EM calorimeter. The different curves correspond
to different selections on the minimum separation between two photons
from the ALP decay applied in the analysis.} \label{fig:results_dralpha}
\end{figure}

\begin{figure}
\centering
\includegraphics[width=0.8\textwidth]{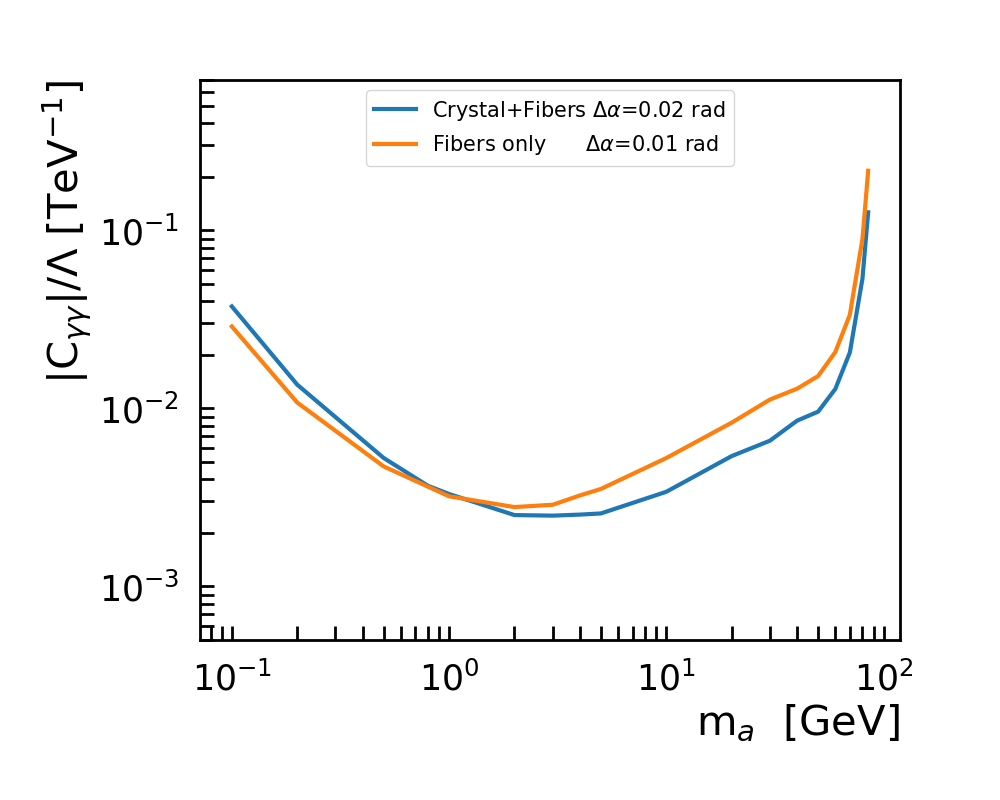}
\caption{95\% CL sensitivity for the $C_{\gamma\gamma}$ coupling 
as a function of $m_{a}$ for the three-photon analysis
for the crystal (blue) and fiber (orange) EM calorimeters.} \label{fig:results_calo}
\end{figure}

%The shape of the distributions display a mild dependence from the test mass,
%differently from lab-frame variables like the angle between $\gamma_1$ and $\gamma_2$,
%or the ratio of their energies.
For the signal, since $a$ is a scalar $\cos\theta_{\gamma_1}$ and $\phi_{\gamma_1}$
have a flat distribution, which is not the case for the background.
For each test mass a signal selection is developed, based on the following steps:
\begin{itemize}
\item
Events with exactly three photons within $|\eta|<2.6$ and with energy $E_{\gamma}>0.1$~GeV 
are selected, and for those events the photons are assigned to the desired ALP kinematics
as described above.
\item
A preselection is applied, requesting that the invariant mass of the three leptons
is above 84.5 (89.5) GeV for the DR calorimeter (crystal calorimeter) respectively.
This cut
%rejects only signal events where PYTHIA  added radiation to the events, 
has approximately 100\% efficiency for signal, and rejects 
possible instrumental backgrounds where a particle may have escaped the detector.
In addition, loose cuts on the ALP mass kinematics and on the angular distance of the
two ALP photons on the face of the calorimeter are applied. These cuts are meant
to remove irrelevant events from the training of the selection
algorithm, but to retain enough Monte Carlo events in the training
and the test samples to ensure a statistically stable selection.
\item
For the sample thus selected, a boosted decision tree (XGBboost \cite{xgboost}) is
trained on five variables, $M_{cut}$, $E_{\gamma_2}/E_{\gamma_1}$,
$\cos\theta_{\gamma_3}$, $\cos\theta_{\gamma_1}$ 
and $\phi_{\gamma_1}$, where the last two variables are in the rest frame
of the ($\gamma_1,\gamma_2$) system.
\item
Finally a cut on the minimal angular distance between $\gamma_1$ and 
$\gamma_2$ in the lab is applied to reject events where the two photons
are expected to be merged in the calorimeter, as explained above.
\item
For a given assumed value of the coupling $C_{\gamma\gamma}$, 
the number of signal events in the test sample is normalised to the 
expected statistics of $6\times10^{12}$ $Z$ bosons, and the QED 3-photon
background to the target luminosity of $205~\mathrm{ab}^{-1}$.
\item
An example of the  XGB probability ($P_{XGB}$) distribution
for signal and background is shown in Figure~\ref{fig:XGB}, for 
a 40 GeV test mass. The final cut on $P_{XGB}$ is 
set at the value optimising the statistical significance $Z$, defined as in \cite{ATLASstat}.
For a given test mass, $Z$ is calculated for a range of values
of $C_{\gamma\gamma}$, and the experimentally accessible region is defined 
as the $C_{\gamma\gamma}/\Lambda$ interval for which $Z\geq2$, corresponding
to a 95\% CL upper limit on the signal strength.
\end{itemize}
The expected reach in the ($C_{\gamma\gamma}, m_a$) plane is
shown in Figure~\ref{fig:results_dralpha}
for the two calorimeter options and for different values of the cut on 
the $\Delta\alpha$ angle. A significant dependence is observed in both cases
for values of $m_{a}$ below 1~GeV.
The performance of the two calorimeter options is compared in Figure~\ref{fig:results_calo},
for the assumed default values of the selection in $\Delta\alpha$.
As expected the crystal calorimeter has a strong advantage where the 
analysis performance is dominated by the photon energy resolution.
The impact point resolution is very similar for the two calorimeter options, 
and the better granularity of the fiber calorimeter is
expected to provide better separation power for two very collimated 
photons. This needs to be confirmed with dedicated full simulation studies 
which are in progress.

%% file: monophoton.tex
The situation in which the ALP decays outside the detector
yields a final state with a monochromatic prompt photon recoiling 
against an invisible particle. This is the monophoton signature, 
which is a standard way of searching for invisible BSM particles 
at $e^+e^-$ colliders.

The specificity of the ALP model resides in  the fact that the photon 
is monochromatic, as the invisible particle is produced in the two-body decay 
of the $Z$  boson, and has an energy very near to the kinematic limit 
of 45.6 GeV see the recoil formula \ref{eq:recoil}, as the values 
of $m_a$ relevant for the analysis are below 1-2~GeV, as shown in 
Figure \ref{fig:lines_alp}.

The $\theta$ coverage of the tracking detector of IDEA extends down 
to 0.15 radians ($|\eta|<2.6$) from the beam line, whereas the calorimeter 
extends down to 0.1 radians ($|\eta|<3$). The experimental topology  is therefore
the request of a photon within $|\eta|<2.6$, and no other particle seen in the
detector.
Two main backgrounds are considered for this analysis:
\begin{itemize}
\item
the irreducible background is the production 
of the photon in association with a $Z$ boson which in turn decays
invisibly into two neutrinos ($\gamma\nu\nu$ background);
\item
the main  reducible background is the associated production of a photon
with two fermions, where both fermions escape the detector.  
The most important of these backgrounds is the
process $e^+e^-\rightarrow\gamma e^+e^-$ which has a very high cross-section 
for topologies where the two electrons are produced at a small angle from 
the beam. However, in the kinematic region corresponding to the signal,
the radiated photon has to be within the angular acceptance of the inner
detector, and carry approximately half of the center-of-mass energy.  
To balance the transverse momentum of the photon, the associated particles
must be produced at a relatively high angle from the beam direction, 
and thus they have very low probability of escaping the detector unseen.
\end{itemize}
In order to separate the signal from the background a preselection is 
applied requiring a single photon with energy in excess of 40~GeV reconstructed
in the calorimeter within $|\eta|<2.6$, no other signal in the calorimeter, 
and no reconstructed tracks in the inner detector. After this selection 
the reducible background  is fully eliminated, and only the $\gamma\nu\nu$ background 
survives the selection.

\begin{figure}
\centering
\includegraphics[width=0.45\textwidth]{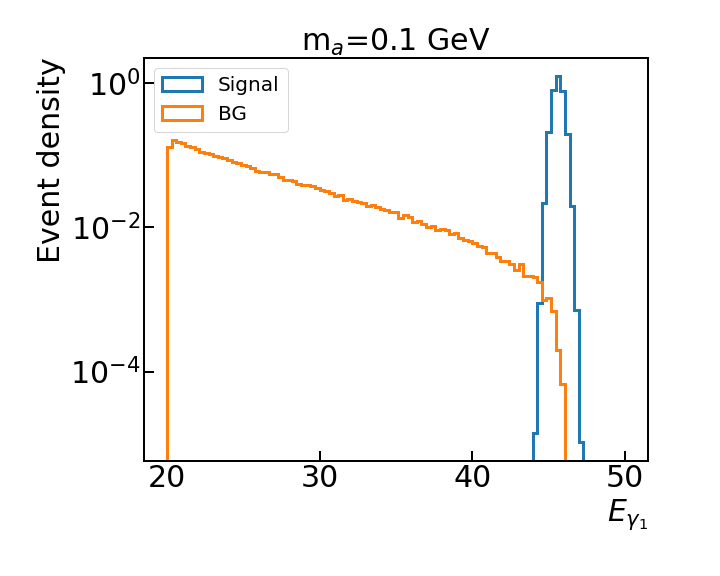}
\includegraphics[width=0.45\textwidth]{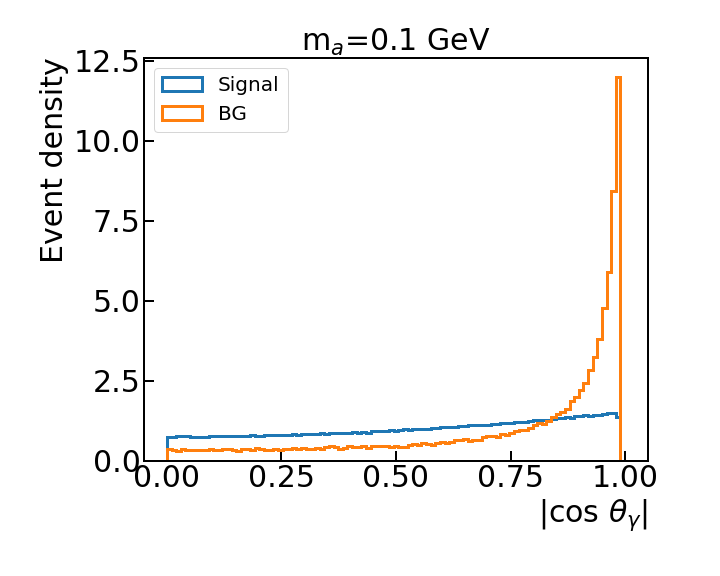}
\caption{Distribution of the variables $E_{\gamma}$ (left) and  $\mathrm{cos}\theta_{\gamma}$ (right) for
signal (blue) and background (orange) for $m_{a}=1$~GeV.} \label{fig:monovar}
\end{figure}

The photon from the $Z$ decay is defined by two variables, its energy, $E_{\gamma}$,
and the cosine of its polar angle $\mathrm{cos}\,\theta_{\gamma}$. The comparison
of signal and background for the two variables is shown in Figure~\ref{fig:monovar}
The already powerful rejection which can be obtained by a threshold on the 
photon energy can be enhanced by a selection on $\mathrm{cos}\,\theta_{\gamma}$.
The two variables, as for the prompt analysis, are combined using a BDT.
A minimum cut on the photon energy of 42 (44.5)~GeV is applied for the fiber 
(crystal) calorimeter respectively before training the tree.
The cut on the output of the tree is chosen to maximise sensitivity 
for each value of $m_{a}$, as explained for the prompt analysis.

\begin{figure}
\centering
\includegraphics[width=0.8\textwidth]{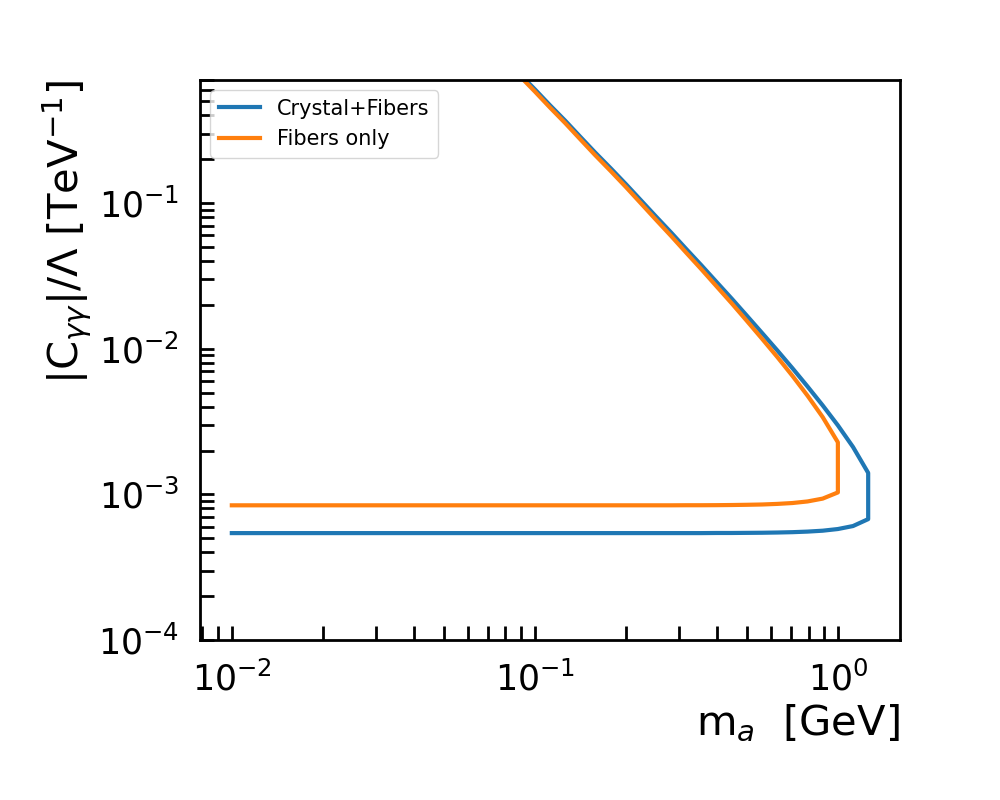}
\caption{Values the $C_{\gamma\gamma}$ coupling for which the monophoton
analysis has a $2\sigma$ sensitivity as a function of $m_{a}$ for the crystal
(blue) and fiber (orange) EM calorimeters.} \label{fig:results_mono}
\end{figure}

The values of the $C_{\gamma\gamma}$ coupling for which this analysis
is sensitive to ALP production at 95\% CL are shown in Figure~\ref{fig:results_mono} 
for the two calorimeter options  as a function of $m_{a}$. 
Given the fact that the background steeply rises with decreasing energy of the photons
the much better resolution of the crystal calorimeter yields a significant
improvement in coverage.

%% file: conclusions.tex
\begin{figure}
\centering
\includegraphics[width=0.9\textwidth]{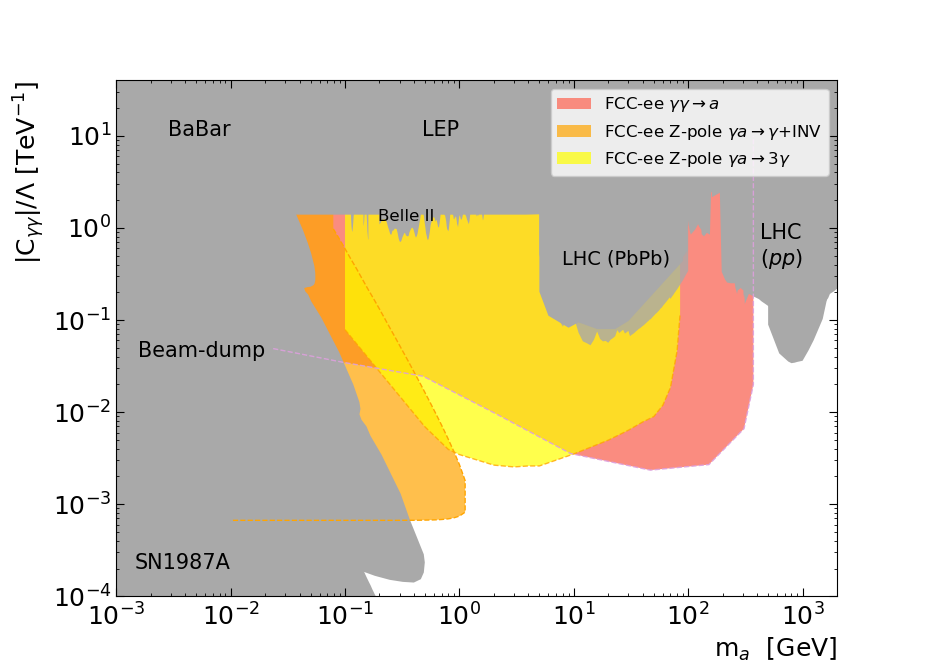}
\caption{Projected sensitivity for ALPs in the photon coupling versus ALP mass plane from $\rm e^+ e^- \to \gamma a \to 3\gamma$ analysis (yellow area) and for the 
$\rm e^+ e^- \to \gamma a \to \gamma + \mathrm{invisible}$ analysis (orange area) at FCC-ee.
The salmon area shows the coverage of  the photon-fusion 
$\gamma\gamma \to  a \to 2\gamma$ process at FCC-ee from \cite{RebelloTeles:2023uig}. 
Other limits are adapted from  Ref.~\cite{Agrawal:2021dbo,Antel:2023hkf} updated to the 
latest LHC results.}\label{fig:results_total}
\end{figure}

The $Z$-pole run at FCC-ee will allow the development of a detailed search 
program for the decay of a $Z$ boson into a photon and an ALP, with 
the ALP subsequently decaying into a pair of photons. Depending on the 
ALP mass and its coupling to the photon and to the $Z$, different search 
strategies have been identified, corresponding to different times of flight 
of the ALP before decaying. A detailed analysis has been performed for 
a three-photon signal, addressing ALPs which decay inside the 
detector, and the monophoton search for ALPs decaying outside 
of the detector. Significant complementary areas of the parameter space 
are covered by the two analyses. The covered area depends on the
energy resolution of the electromagnetic calorimeter, on the resolution 
on the measurement of the impact position of the photon on the calorimeter,
and on how well the invariant mass can be reconstructed when the two 
photons from the ALP decays are well collimated. Reach curves as a function 
of these performance parameters have been obtained,

The results of this study are shown in Figure~\ref{fig:results_total}, together 
with the results from \cite{RebelloTeles:2023uig}, compared to existing ALP limits from 
 Ref.~\cite{Agrawal:2021dbo,Antel:2023hkf} updated to the latest LHC results.
The three-photon analysis would extend significantly the existing LHC limits 
in the mass region 1-90 GeV. The monophoton analysis covers the region below 0.1 and 1
GeV which is beyond the reach of the beam dump experiments, and the \mbox{$\gamma\gamma\rightarrow a$} of Ref.~\cite{RebelloTeles:2023uig} would extend the mass coverage
to $m_a\sim350$~GeV for comparable values of $C_{\gamma\gamma}$. Thanks to the FCC-ee
the complete mass range up to $\sim350$~GeV will be explored for couplings larger than
a few $10^{-3}$ TeV$^{-1}$.